\documentclass[12pt]{article}
\usepackage{amsfonts}
\usepackage{amssymb}
\usepackage{graphics,psboxit,amsmath}
\usepackage{subfigure}
\usepackage{graphicx}
\usepackage{verbatim}
\usepackage{epic}

\def\hybrid{\topmargin 0pt \oddsidemargin 0pt 
        \headheight 0pt \headsep 0pt
        \textwidth 16,5cm 
        \textheight 23cm 
        \marginparwidth .875in
        \parskip 5pt plus 1pt \jot = 1.5ex}

\hybrid

\catcode`\@=11

\def\marginnote#1{}
\newcount\hour
\newcount\minute
\newtoks\amorpm
\hour=\time\divide\hour by60
\minute=\time{\multiply\hour by60 \global\advance\minute by-\hour}
\edef\standardtime{{\ifnum\hour<12 \global\amorpm={am}%
        \else\global\amorpm={pm}\advance\hour by-12 \fi
        \ifnum\hour=0 \hour=12 \fi
        \number\hour:\ifnum\minute<10 0\fi\number\minute\the\amorpm}}
\edef\militarytime{\number\hour:\ifnum\minute<10 0\fi\number\minute}

\def\draftlabel#1{{\@bsphack\if@filesw {\let\thepage\relax
   \xdef\@gtempa{\write\@auxout{\string
      \newlabel{#1}{{\@currentlabel}{\thepage}}}}}\@gtempa
   \if@nobreak \ifvmode\nobreak\fi\fi\fi\@esphack}
        \gdef\@eqnlabel{#1}}
\def\@eqnlabel{}
\def\@vacuum{}
\def\draftmarginnote#1{\marginpar{\raggedright\scriptsize\tt#1}}

\def\draft{\oddsidemargin -.5truein
        \def\@oddfoot{\sl preliminary draft \hfil
        \rm\thepage\hfil\sl\today\quad\militarytime}
        \let\@evenfoot\@oddfoot \overfullrule 3pt
        \let\label=\draftlabel
        \let\marginnote=\draftmarginnote
   \def\@eqnnum{(\theequation)\rlap{\kern\marginparsep\tt\@eqnlabel}%
\global\let\@eqnlabel\@vacuum} }

\def\draft2{
        \def\@oddfoot{\sl preliminary draft \hfil
        \rm\thepage\hfil\sl\today\quad\militarytime}
        \let\@evenfoot\@oddfoot \overfullrule 3pt
        \let\label=\draftlabel
        \let\marginnote=\draftmarginnote
   \def\@eqnnum{(\theequation)\rlap{\kern\marginparsep\tt\@eqnlabel}%
\global\let\@eqnlabel\@vacuum} }

\def\preprint{\twocolumn\sloppy\flushbottom\parindent 2em
        \leftmargini 2em\leftmarginv .5em\leftmarginvi .5em
        \oddsidemargin -.5in \evensidemargin -.5in
        \columnsep .4in \footheight 0pt
        \textwidth 10.in \topmargin -.4in
        \headheight 12pt \topskip .4in
        \textheight 6.9in \footskip 0pt
        \def\@oddhead{\thepage\hfil\addtocounter{page}{1}\thepage}
        \let\@evenhead\@oddhead \def\@oddfoot{} \def\@evenfoot{} }

\def\numberbysection{\@addtoreset{equation}{section}
        \def\theequation{\thesection.\arabic{equation}}}

\def\underline#1{\relax\ifmmode\@@underline#1\else
        $\@@underline{\hbox{#1}}$\relax\fi}

\def\titlepage{\@restonecolfalse\if@twocolumn\@restonecoltrue\onecolumn
     \else \newpage \fi \thispagestyle{empty}\c@page\z@
        \def\thefootnote{\fnsymbol{footnote}} }

\def\endtitlepage{\if@restonecol\twocolumn \else \newpage \fi
        \def\thefootnote{\arabic{footnote}}
        \setcounter{footnote}{0}} 

\catcode`@=12
\relax

\def\figcap{\section*{Figure Captions\markboth
        {FIGURECAPTIONS}{FIGURECAPTIONS}}\list
        {Figure \arabic{enumi}:\hfill}{\settowidth\labelwidth{Figure
999:}
        \leftmargin\labelwidth
        \advance\leftmargin\labelsep\usecounter{enumi}}}
 \relax
\def\tablecap{\section*{Table Captions\markboth
        {TABLECAPTIONS}{TABLECAPTIONS}}\list
        {Table \arabic{enumi}:\hfill}{\settowidth\labelwidth{Table
999:}
        \leftmargin\labelwidth
        \advance\leftmargin\labelsep\usecounter{enumi}}}
 \relax
\def\reflist{\section*{References\markboth
        {REFLIST}{REFLIST}}\list
        {[\arabic{enumi}]\hfill}{\settowidth\labelwidth{[999]}
        \leftmargin\labelwidth
        \advance\leftmargin\labelsep\usecounter{enumi}}}
 \relax
\makeatletter
\newcounter{pubctr}
\def\publist{\@ifnextchar[{\@publist}{\@@publist}}
\def\@publist[#1]{\list
        {[\arabic{pubctr}]\hfill}{\settowidth\labelwidth{[999]}
        \leftmargin\labelwidth
        \advance\leftmargin\labelsep
        \@nmbrlisttrue\def\@listctr{pubctr}
        \setcounter{pubctr}{#1}\addtocounter{pubctr}{-1}}}
\def\@@publist{\list
        {[\arabic{pubctr}]\hfill}{\settowidth\labelwidth{[999]}
        \leftmargin\labelwidth
        \advance\leftmargin\labelsep
        \@nmbrlisttrue\def\@listctr{pubctr}}}
 \relax
\makeatother

\def\ba{\begin{equation}}
\def\ea{\end{equation}}

\def\e{\epsilon}

\def\no{\noindent}

\def\IR{\relax{\rm I\kern-.18em R}}

\begin{document}


\renewcommand{\theequation}{\thesection.\arabic{equation}}
\csname @addtoreset\endcsname{equation}{section}

\newcommand{\eqn}[1]{(\ref{#1})}
\newcommand{\be}{\begin{eqnarray}}
\newcommand{\ee}{\end{eqnarray}}
\newcommand{\non}{\nonumber}
\begin{titlepage}
\strut\hfill
\vskip 1.3cm
\begin{center}


{\large \bf Contractions of quantum algebraic structures}
\footnote{{\tt Proceedings contribution to the ``9th Hellenic School on
Elementary Particle Physics and Gravity'' Corfu, September 2009. Based on a talk given by A.D.}}

\vskip 0.5in

{\bf Anastasia Doikou}\phantom{x} and\phantom{x} {\bf Konstadinos Sfetsos}
\vskip 0.1in

Department of Engineering Sciences, University of Patras,\\
26110 Patras, Greece\\

\vskip .1in

\vskip .15in

{\footnotesize {\tt adoikou@upatras.gr},
\ \ {\tt sfetsos@upatras.gr}}\\

\end{center}

\vskip .8in

\centerline{\bf Abstract}

\no
A general framework for obtaining certain
types of contracted and centrally extended algebras is presented.
The whole process relies on the existence of quadratic algebras,
which appear in the context of boundary integrable models.

\vfill
\no


\end{titlepage}
\vfill
\eject



\def\baselinestretch{1.0}
\baselineskip 15 pt
\no

\section{Introduction}

We show that the symmetry breaking mechanism due to the presence
of appropriate boundary conditions may be
exploited in order to obtain centrally extended algebras via suitable contraction procedures.
We use the boundary algebra to obtain the relevant Casimir operator. We explicitly demonstrate that
this is perhaps the simplest and most straightforward way
to obtain the Casimir operator of usual and
deformed Lie algebras.

\noindent
One main point of this investigation is that
we are able to show that the associated open transfer matrix commutes
with the elements of the emerging contracted algebra.
We study here the simplest case, that is the $E_{2}^c$ algebra,
in order to illustrate the procedure followed, however this description
may be generalized to more complicated algebraic structures.

\noindent
This brief note is based on \cite{doikousfetsos}
where the interested reader can find all details of the construction.

\section{Quadratic algebras}

We give first a short review of the fundamental quadratic algebraic relations,
ruling quantum integrable models, that is the
Yang--Baxter and reflection equations. The Yang--Baxter equation \cite{baxter} is defined as
\be
R_{12}(\lambda_{1} -\lambda_{2})\ R_{13}(\lambda_{1})\
R_{23}(\lambda_{2})\ =\ R_{23}(\lambda_{2})\ R_{13}(\lambda_{1})\
R_{12}(\lambda_{1}-\lambda_{2})\ ,
\label{ybe2}
\ee
acting on
${\mathbb V}^{\otimes 3}$ and $R \in \mbox{End}({\mathbb
V}^{\otimes 2})$ $~R_{12} = R \otimes {\mathbb I}$, $R_{23} =
{\mathbb I} \otimes R$, where $R$ physically describes
the scattering among particle-like excitations
displayed in integrable models. Given an $R$ matrix
we introduce the following fundamental
algebraic relations \cite{tak}, which  define the
algebra ${\cal A}$ (see e.g. \cite{tak, fad})
\be
R_{12}(\lambda_1-\lambda_2)\ L_1(\lambda_1)\ L_2(\lambda_2)\ =\
L_2(\lambda_2)\ L_1(\lambda_1)\ R_{12}(\lambda_1-\lambda_2)\ ,
\label{fundam}
\ee
where $L \in \mbox{End}({\mathbb V}) \otimes
{\cal A}$. This allows the construction of tensorial
representations of the later algebra as \cite{tak, fad}
\be
T_a(\lambda)  =
L_{aN}(\lambda-\theta_N)\ L_{a N-1}(\lambda-\theta_{N-1}) \ldots
L_{a2}(\lambda-\theta_2)\ L_{a1}(\lambda-\theta_1)\ ,
\label{mono}
\ee
where $T(\lambda) \in \mbox{End}({\mathbb V}) \otimes {\cal
A}^{\otimes N}$. Traditionally, the space $a$ is called
``auxiliary", whereas the spaces $1, \dots, N$ are called
``quantum''. For simplicity we usually suppress all quantum spaces
when writing down the monodromy matrix. The free to choose
complex parameters $\theta_i$ are called inhomogeneities. Using the
fundamental algebra (\ref{fundam}) one may show that
\be
\Big [ tr
T(\lambda),\ tr T(\mu) \Big ] =0\ ,
\ee
where $tr T(\lambda) \in
{\cal A}^{\otimes N}$ and the trace is taken over the
auxiliary space $a$. The latter relation guarantees the integrability of the
system.

\noindent
We next introduce the reflection equation associated to the reflection algebra ${\cal R}$.
It is given by \cite{cherednik, sklyanin}
\be
R_{12}(\lambda_{1} -\lambda_{2})\ {\mathbb K}_{1}(\lambda_{1})\ R_{21}(\lambda_{1}
+\lambda_{2})\ {\mathbb K}_{2}(\lambda_{2})=
{\mathbb K}_{2}(\lambda_{2})\ R_{12}(\lambda_{1} +\lambda_{2})\
 {\mathbb K}_{1}(\lambda_{1})\ R_{21}(\lambda_{1} -\lambda_{2})\ ,
\label{re2}
\ee
acting on ${\mathbb V}^{\otimes 2} $ and as customary, we follow the notation
${\mathbb K}_{1} ={\mathbb K} \otimes {\mathbb I}$ and $ {\mathbb K}_{2} =
{\mathbb I} \otimes {\mathbb K}$.
Also $R_{21} = {\cal P}\ R_{12}\ {\cal P}$, where ${\cal P}$ is the permutation operator,
acting as ${\cal P} (a \otimes b)= b \otimes a$ and
in addition ${\mathbb K} \in \mbox{End}({\mathbb V})\otimes {\cal R}$.
In general, the representations of the later algebra may be expressed
as \cite{sklyanin}
\be
{\mathbb K}(\lambda) = L(\lambda)\ K(\lambda)\otimes {\mathbb I}\ L^{-1}(-\lambda)\ ,
\label{kk1}
\ee
where the matrix $K$ is a $c$-number representation of
the aforementioned algebra, called also the reflection matrix and physically describes the interaction
of a particle-like excitation
with the boundary. Tensor
representations of these algebra are given by (\ref{kk1}) after replacing $L$
by $T$ defined in (\ref{mono}).
We define the open transfer matrix as \cite{sklyanin}
\be
t(\lambda) = tr\{K^+(\lambda) {\mathbb K}(\lambda)\}\ ,
\label{transfer}
\ee
where $K^+$ matrix is a $c$-number solution of the reflection algebras.
With the help of the quadratic exchange relations one
may show that (see e.g. \cite{sklyanin})
\be
[t(\lambda),\ t(\mu) ] =0\ ,
\label{coons}
\ee
which again guarantees the integrability of the system under consideration.

\section{The $E_{2}^c$ extended algebra }

We aim at constructing the centrally extended $E_{2}^c$ algebra. To
achieve this we start from the $gl_3$ spin chain and break the
symmetry down to $sl_2 \otimes u(1)$ by using the results and techniques
developed in \cite{done}. Indeed, by implementing appropriate boundary
conditions one can break the $gl_n$ symmetry of a spin chain model
to $gl_l \otimes gl_{n-l}$, where $l$ is an integer depending on the
choice of boundary. We will exploit this phenomenon in order to
perform a contraction of the boundary algebra to $E_{2}^c$.

\noindent
The $gl_n$ algebra is generated by $J^{+(k)},\ J^{-(k)}$ and $e^{(i)}$, with $i=1,2,\dots , n$.
Defining
$\displaystyle s^{(k)} = e^{(k)} -e^{(k+1)}$ they
satisfy the commutation relations
\be
[J^{+(k)},\ J^{-(l)}] = \delta_{kl} s^{(k)}\ , \qquad
[s^{(k)},\ J^{\pm (l)}] =\pm (2 \delta_{kl} - \delta_{k\ l+1} - \delta_{k\ l-1}) J^{\pm (l)}\
\label{sllk}
\ee
and $\sum_{i=1}^n{e^{(i)}}$ belongs to the center of the algebra.
We will focus here to the $gl_3$ algebra.
The $L$ matrix is expressed as
$L(\lambda) = \lambda + i {\mathbb P}$,
where ${\mathbb P}$ in terms of the $gl_3$ elements takes the form
\be
{\mathbb P} = \left(
\begin{array}{ccc}
e^{(1)} &J^{-(1)} &\Lambda^{+}\\
J^{+(1)} &e^{(2)} &J^{-(2)} \\
\Lambda^{-} &J^{+(2)} &e^{(3)} \\ \end{array} \right )\ ,\qquad
\mbox{where} \qquad  \Lambda^{\pm} = \pm[J^{\pm(1)},\ J^{\pm(2)} ]\ .
\label{def3}
\ee
We choose as $K$ in \eqn{kk1} the following diagonal matrix
\be
K(\lambda) =k=\mbox{diag}(1,\ 1,\ -1) \
\label{dk}
\ee
and expand
${\mathbb K}(\lambda)$ as
\be {\mathbb K}_0(\lambda) =
L_{01}(\lambda)\ k_0\ L^{-1}_{01}(-\lambda) = k +
\sum_{j}{{\mathbb K}^{(j-1)} \over \lambda^{j}}\ .
\ee
The first two coefficients are
\be {\mathbb K}^{(0)}  = i
\Big ( {\mathbb P}_{01}k_0 + k_0 {\mathbb P}_{01} \Big )\ ,
\qquad {\mathbb K}^{(1)} =  - {\mathbb P}_{01}\ k_0\ {\mathbb P}_{01} - k_0
 {\mathbb P}_{01}^2 \ .
\label{first}
\ee
Clearly, what remains are the generators of the $sl_2 \otimes u(1)$ algebra. Specifically,
$\big(\displaystyle J^{\pm(1)},\  s^{(1)}\big)$
satisfy the $sl_2$ commutation relations, whereas $c=\sum_i e_i^{(3)}$
commutes with everything.
The first two conserved quantities are given by taking the trace over the auxiliary space
in ${\mathbb K}^{(0)},\ {\mathbb K}^{(1)}$.
If for notational convenience, we set
\be
s^{(1)}\equiv 2 J\ , \qquad  e^{(3)} \equiv 2 \tilde J\ , \qquad  J^{-(1)} \equiv -J^-\ ,
\qquad J^{+(1)}\equiv J^+\ .
\ee
and also consider the transfer matrix expansion as
\be
t(\lambda) = \sum_k {t^{(k-1)}\over \lambda^k}\ ,
\label{tt}
\ee
we obtain the integrals of motion
\be
t^{(0)} =  c - 4 \tilde J,
\qquad t^{(1)} \propto J^2 -{1\over 2}\{J^{+},\ J^{-}\} -\tilde J^2 -c\tilde J +{c^2\over 4}\ .
\ee
From the first integral of
motion it is clear that $e^{(3)} = 2 \tilde J$ is also a conserved quantity,
so that in the second charge we may drop the last two terms. In conclusion, the conserved changes
can be taken to be
\be
I^{(0)} = c - 4 \tilde J\ , \qquad  I^{(1)}= J^2
-{1\over 2}\{J^{+},\ J^{-}\} -\tilde J^2\  . 
\label{1part}
\ee
Similarly, the $N$-tensor representations are obtained
in a straightforward manner (see \cite{doikousfetsos} for more details).

\noindent
We will exploit the breaking of the symmetry due to the presence of non-trivial
integrable boundaries to obtain the centrally
extended $E_{2}^c$ algebra.
We consider a contraction known in the mathematics
literature as a Saletan contraction and
is distinct from the more common Inon\"u--Winger contraction.
It is closely related to the so called Penrose limit in gravity,
that constructs a plane
wave starting from any gravitational
background by magnifying the region around a null geodesic. This was
first used in the string literature in WZW models \cite{sfetsos},
more recently in various supersymmetric brane solutions
of string and M-theory \cite{BFHPPenrose} and has been instrumental
in understanding issues within the AdS/CFT correspondence involving sectors of large quantum
numbers \cite{BMN}. According to this contraction
\be
J^{\pm} = {1 \over \sqrt{2\epsilon}} P^{\pm}\ , \qquad  J ={1 \over 2}\left(T+{F\over \epsilon}\right)\ ,
\qquad  \tilde J = - {F \over 2 \epsilon}\ ,\qquad  \e\to 0\ .
\label{map}
\ee
Then one obtains the following commutation relations that define the $E_{2}^c$ algebra
\be
[P^+,\ P^- ] = -2F\ , \qquad  [T,\ P^{\pm}] = \pm P^{\pm}\ ,
\label{ec2}
\ee
where $F$ is an {\it exact} central element of the algebra.
It is obvious that the conserved quantities,
after contracting and keeping the leading order contribution are
\be
I^{(0)} = F\ , \qquad I^{(1)} = T F-{1
\over 2} \{P^+,\ P^- \}\ .
\label{part1}
\ee
Recall the representation of the reflection algebra
\be
{\mathbb K}(\lambda) & = &  L(\lambda)\ k\ L^{-1}(-\lambda) = (1 +{i\over \lambda}{\mathbb P})\ k\
(1 +{i\over \lambda}{\mathbb P} -{1\over \lambda^2}{\mathbb P}^2
-{i \over \lambda^3}{\mathbb P}^3 + {1\over \lambda^4}{\mathbb P}^4 \ldots)
\nonumber\\
& = & k + {i\over \lambda}({\mathbb P} k + k {\mathbb P}) -{1 \over \lambda^2}({\mathbb P} k {\mathbb P} +
k{\mathbb P}^2) - {1\over \lambda^2} ({\mathbb P}k {\mathbb P}^2 + k {\mathbb P}^3 ) \ldots\ ,
\label{exp11}
\ee
where $k$ is given in (\ref{dk}). Then, after taking the trace over the auxiliary space
and recalling (\ref{tt}), we obtain
\be
t^{(k-1)} \propto \sum_{a, b}({\mathbb P}_{ab}\ k_{bb}\ {\mathbb P}^{k-1}_{ba} + k_{aa}\
{\mathbb P}_{aa}^k )\ .
\label{tk}
\ee
This is a quite general result that holds irrespectively of the particular application
based on the $sl_2\times u(1)$ algebra we have just employed. Indeed, based on this example
we may infer that, before the contraction, $t^{(k)}$ are the higher Casimir operators
of $sl_n \otimes u(2)$. After
the contraction is taken one has to consistently keep only the highest order
contribution in the ${1 \over \epsilon}$ expansion of each $t^{(k)}$. Then, each one of $t^{(k)}$
commutes by construction with the contracted algebra ($E_2^c$ in our elementary example)
and clearly, the same does the transfer matrix. This logic may be generalized
to contractions of any higher rank algebras and similarly for generic $N$-particle
representations (see \cite{doikousfetsos}). In fact, this argument
holds independently of the context one realizes
the contraction (see e.g. \cite{sfetsos}). More precisely, having in general a
set of Casimir operators of say the $gl_n$ algebra,
after the contraction is taken one consistently should keep the highest order
contribution in order to obtain the
contracted Casimir quantities. Depending on the rank of the considered algebra
the expansion of $t(\lambda)$ should truncate at some point,
or in other words the higher
Casimir quantities should be trivial combinations of the lower ones.

\noindent
To simplify
the analysis and clearly demonstrate the main ideas we restricted our discussion to examples involving
originally $gl_3$ symmetry. A similar construction for the $q$ deformed case has been also
analyzed in \cite{doikousfetsos}.
A natural extension of the present work is to consider generic symmetry breaking of the type
${\mathrm G} \otimes {\mathrm H}$ where ${\mathrm G}$ and
${\mathrm H}$ are generic algebras (${\mathrm H} \subset {\mathrm G}$),
and then follow a contraction procedure similar to the ones for ordinary Lie and current algebras
\cite{sfetsos}.

\noindent
Finally, we note that our systematic approach for taking limits in general algebraic structures has
been instrumental in resolving,
in a natural way, a quite old misunderstanding concerning the opposite to contraction
procedure, commonly known as expansion. For the details the interested reader is referred to
\cite{doikousfetsos}.

\end{document}